\documentclass[12pt]{article}
\title{ From 3 nucleons to 3 quarks}
\author{  Yu.A.Simonov}
\date{}
\newcommand{\beq}{\begin{eqnarray}}
 \newcommand{\eeq}{\end{eqnarray}}
\newcommand{\be}{\begin{equation}}
 \newcommand{\ee}{\end{equation}}

 \def\la{\mathrel{\mathpalette\fun <}}
\def\ga{\mathrel{\mathpalette\fun >}}
\def\fun#1#2{\lower3.6pt\vbox{\baselineskip0pt\lineskip.9pt
\ialign{$\mathsurround=0pt#1\hfil ##\hfil$\crcr#2\crcr\sim\crcr}}}
\newcommand{\veX}{\mbox{\boldmath${\rm X}$}}
\newcommand{{\SD}}{\rm SD}
\newcommand{\pp}{\prime\prime}
\newcommand{\veY}{\mbox{\boldmath${\rm Y}$}}
\newcommand{\vex}{\mbox{\boldmath${\rm x}$}}
\newcommand{\vey}{\mbox{\boldmath${\rm y}$}}
\newcommand{\ver}{\mbox{\boldmath${\rm r}$}}
\newcommand{\vesig}{\mbox{\boldmath${\rm \sigma}$}}
\newcommand{\vedelta}{\mbox{\boldmath${\rm \delta}$}}
\newcommand{\veP}{\mbox{\boldmath${\rm P}$}}
\newcommand{\vep}{\mbox{\boldmath${\rm p}$}}
\newcommand{\veq}{\mbox{\boldmath${\rm q}$}}
\newcommand{\vez}{\mbox{\boldmath${\rm z}$}}
\newcommand{\veS}{\mbox{\boldmath${\rm S}$}}
\newcommand{\veL}{\mbox{\boldmath${\rm L}$}}
\newcommand{\veR}{\mbox{\boldmath${\rm R}$}}
\newcommand{\ves}{\mbox{\boldmath${\rm s}$}}
\newcommand{\vek}{\mbox{\boldmath${\rm k}$}}
\newcommand{\ven}{\mbox{\boldmath${\rm n}$}}
\newcommand{\veu}{\mbox{\boldmath${\rm u}$}}
\newcommand{\vev}{\mbox{\boldmath${\rm v}$}}
\newcommand{\veh}{\mbox{\boldmath${\rm h}$}}
\newcommand{\verho}{\mbox{\boldmath${\rm \rho}$}}
\newcommand{\vexi}{\mbox{\boldmath${\rm \xi}$}}
\newcommand{\veta}{\mbox{\boldmath${\rm \eta}$}}
\newcommand{\veB}{\mbox{\boldmath${\rm B}$}}
\newcommand{\veH}{\mbox{\boldmath${\rm H}$}}
\newcommand{\veE}{\mbox{\boldmath${\rm E}$}}
\newcommand{\veJ}{\mbox{\boldmath${\rm J}$}}
\newcommand{\veal}{\mbox{\boldmath${\rm \alpha}$}}
\newcommand{\vegam}{\mbox{\boldmath${\rm \gamma}$}}
\newcommand{\vepar}{\mbox{\boldmath${\rm \partial}$}}
\newcommand{\llan}{\langle\langle}
\newcommand{\rran}{\rangle\rangle}
\newcommand{\lan}{\langle}
\newcommand{\ran}{\rangle}
\begin{document}
\maketitle

\begin{abstract}
Some short history of few-body methods originated from the famous
Skornyakov-Ter-Martirosyan equation is given, including latest
development of Faddeev formalism and Efimov states. The 3q system
is shown to require an alternative, which is provided by the
hyperspherical method ($K$-harmonics) which is  highly successful
for baryons.
\end{abstract}

\section{Introduction}

The Skornyakov and Ter-Martirosyan paper \cite{1} which appeared
in 1956 marked the beginning of  a new era in  few-body physics,
when a somewhat neglected part of nuclear physics was promoted to
the successful domain of theoretical physics. As a result the
few-body science has become a field accumulating fast developing
methods: L.D.Faddeev generalized the Skornyakov-Ter-Martirosyan
Equation (STME) \cite{2} and has given a rigorous mathematical
foundation for  the theory of 3 particles \cite{3}, many numerical
methods have been introduced, for a review and references  see
\cite{4}. As an immediate consequence of STME, a new effect was
found in 1971, called the Efimov effect \cite{5}, which is studied
till now with respect to possible experimental consequences
\cite{6}.

The STME and Faddeev technic is most useful when particles are
nearly on-shell, so that e.g. 3-body results do not depend much on
the potential shape, but rather described by the on-shell two-body
$t$-matrix as it is for the quartet $n-d$ scattering. The bound
states of tritium and $^3He$ provide another example, where the
interaction at small distances (far off- shell) is important. To
treat such systems an alternative method - the  Hyperspherical
Formalism (HF) (or $K$-harmonics method) was developed  and the
system of  the Schroedinger-like equations  was written \cite{7}.
Its development was marked with many successful applications both
in nuclear and atomic physical see e.g. \cite{8,9}. Recently it
was understood that HF is probably the best suitable for systems
with confinement such as 3 quarks, where the interaction is a
three-body one,  and confining so that the $t$-matrix formalism
cannot be applied. Accuracy of HF as applied to the 3q system was
found to be remarkably good \cite{10,11} allowing for the 1\% bias
in the baryon mass \cite{12}.

This talk is intended to demonstrate the physics of the 3-body
system, and a qualitative analysis of two alternative approaches
discussed above.

\section{The STME and Faddeev approach}

In  the system of 3 equal-mass particles with arbitrary numeration
one can introduce the total kinetic energy $E$ and the momentum
$\vek$ in the pair (\ref{2},\ref{3}) and the relative momentum
$\vep$ of particle 1, namely $\vep
=\frac{\vek_2+\vek_3}{3}-\frac23 \vek_1,~~
\vek=\frac{\vek_2-\vek_3}{2}$.

The symmetric function of the ground state $\Psi_{symm}$ is
expressed through partial w.v.
\be
\Psi_{symm}=\psi(\vek_{23},\vep_1) +\psi(\vek_{31}, \vep_2) + \psi
(\vek_{12}, \vep_3)\label{1}\ee with the normalization condition
\be \int |\Psi_{symm} |^2 d^3\vek d^3 \vep =1.\label{2}\ee It is
convenient to extract the free 3-body Green's function,
introducing
\be
\psi(\vek, \vep) =\frac{\chi(\vek, \vep)}{\vek^2+\frac34 \vep^2
-mE}\label{3}\ee and the 3-body rescattering equation, equivalent
to the summing the "bridge" Feynman diagrams (nonrelativistic) is
\cite{2}
\be
\chi (\vek,\vep) =\chi_0(\vek,\vep) -2 \int m
\frac{t(k,|\frac{\vep}{2} +\vep'| , E-\frac34
\frac{p^2}{m})\chi(|\vep +\frac{p'}{2}|, p')d\vep'}{\vep^{\prime
2}+ \vep\vep' +\vep^2 -mE}.\label{4}\ee Here $t(k,k',\varepsilon)$
is the 2-body $t$-matrix, representing the "knot" in a bridge
diagram, and
\be
\chi_0 (\vek,\vep) = -2  m \frac{t(k,\frac{\vep}{2} +\vep_0,
E-\frac34 \frac{p^2}{m})\varphi_\alpha(\vep +\frac12\vep_0)}{\vep^
2+ \vep^2_0+\vep\vep_0-mE},\label{5}\ee where $\varphi_\alpha$ is
the 2-body bound state, while $\vep_0$ is the momentum of incident
particle.

Near the   bound-state pole $t$-matrix can be written as
\be
t(k,k',\varepsilon) = \frac{g(k\varepsilon)
g(k',\varepsilon)}{(2\pi)^2 m (\alpha+ i
\sqrt{2m\varepsilon})}+O(r_0)\label{6}\ee where $\alpha=1/a, a$ is
the scattering length and $g(k,\varepsilon)$ formfactor,
$g(0,0)=1$ and  $g(k,\varepsilon)$ fast decreases when $k\sim
1/r_0$ and $\varepsilon \sim \frac{1}{mr_0^2}$.

Let us assume now that the range of integration in (\ref{4}) is
small $p, p'\ll 1/r_0$. Then one can  insert (\ref{6}) in
(\ref{4}) with  $g\cong 1$, and one  gets- for the 3-body
bound-state w.f.
\be
(\alpha - \sqrt{\frac34\vep^2-E}) \chi (\vek, \vep) + 8\pi \int
\frac{d\vep'}{(2\pi)^3}
\frac{\chi(|\vep+\frac{\vep'}{2}|,p')}{\vep^{\prime 2} +\vep \vep'
+ \vep^2 - mE} =0.\label{7}\ee

This is the STME for a 3-body bound state. The off-shell
generalization   of STME is the Faddeev equation(\ref{4}). As it
was correctly stated in \cite{1}, the bound-state equation
(\ref{7}) cannot be used  for tritium and $^3He$, since it has no
lower bound for energy due to the Thomas theorem \cite{13}. This
can be easily understood rewriting (\ref{7}) in the form $\chi =
\int K\chi d\vep'$, and calculating the norm of $K, \parallel K
\parallel = \int d\vep d\vep (K(\vep, \vep'))^2$, which diverges
logarithmically at large momenta, implying that there are formally
infinitely many bound states. The physical situation corresponds
to the cut-off  form-factors $g(k,\varepsilon)$ present in $K$,
which leads to the finite result for the norm $\parallel K
\parallel$.

A specific situation occurs when the 2-body scattering length $a$
is  large, $a \gg r_0$. Then the number of bound states lying
between $\left( -\frac{1}{ma^2}\right)$ and $\left(
-\frac{1}{mr^2_0}\right)$ is approximately equal to
\be
N\sim \frac{1}{\pi} \ln \frac{|a|}{r_0}\label{8}\ee
 and when $|a|$  is increasing, $|a|\to \infty$, there appears an
 accumulation point of bound states (the Efimov effect\cite{5}).
For 3 nucleons however $N<1$ and the effect is absent, but for
three $^4He$  atoms $a=104 \AA, r_0\cong 7 \AA$ and the effect is
theoretically possible \cite{6}.

Since the Efimov states are almost on-shell, it is convenient to
calculate them using the 3-body unitarity and the $N/D$ method
\cite{14}. Numerical results obtained (see Fig. 7 of \cite{14})
support the estimate (\ref{8}) and yield the explicit position of
levels near the energy threshold.

To conclude with the bound state equation (\ref{7}) it is
interesting to study the properties of the bound wave function,
e.g. the size of the bound system. Here one encounters an
important difference between 2- and 3-body systems \cite{15}.
Namely the 2-body loosely bound system with a small binding energy
$\varepsilon$, $m\varepsilon r^2_0\ll 1 $ has a radius of the
order of $r_2=\frac{1}{\sqrt{m\varepsilon}}, r_2\gg r_0$. For the
3-body system the situation may be twofold. In case when a  bound
2-body system exists as a subsystem and the 3-body bound state is
close to the 2+1 threshold, one has a quasi-two-body situation,
whereas when 2-body bound subsystems are absent the size of the
3-body bound state is always $r_0$ however small binding energy is
\cite{16}. Modern calculations of 3-body bound states in the
framework of STME and its development  -- Faddeev equations are
done for $^3H$ and $^3 He$ using systems of around 30 equations
and  exploiting realistic potentials describing $NN$ scattering
and bound states in the large energy interval (0-350 MeV). see
e.g. the review in \cite{17}. Unfortunately results of
calculations yield significant underbinding of around 10-15\%, may
be due to three-body forces, which are not exactly known, hence
final results are model dependent, see \cite{18} for an example
and references.

We go now to the $dN$ scattering which was also the topic of the
primary paper \cite{1}. The corresponding equations look like
\be
\begin{array}{c}\frac{(\sqrt{3k^2/4-ME}-\alpha_t)}{k^2-k^2_0} a_{3/2}(\vek,
\vek_0) =\frac{-1}{k^2_0+k^2+\vek\vek_0-ME}- \\ -\int \frac{4\pi
a_{3/2} (\vek', \vek_0)}{(k^2+k^2+\vek \vek'-ME)(k^2-k^2_0)}
\frac{d\vek'}{(2\pi)^3}; \\
 \frac{(\sqrt
{3k^2/4-ME}-\alpha_t)}{k^2-k^2_0} a_{3/2}(\vek, \vek_0)=
\frac{1/2}{k^2_0+k^2+\vek\vek_0-ME}+\\ +\int \frac{4\pi \{1/2
a_{1/2} (\vek', \vek_0)+3/2 b_{1/2} (\vek',
\vek_0)\}}{(k^2+k^2+\vek \vek'-ME)(k^2-k^2_0)}
\frac{d\vek'}{(2\pi)^3},\\
\frac{(\sqrt{3k^2/4-ME}-\alpha_t)}{k^2-k^2_0}b_{1/2} (\vek,
\vek_0) =\frac{3/2}{k^2_0+k^2+\vek\vek_0-ME}+\\ +\int \frac{4\pi
\{3/2 a_{1/2} (\vek', \vek_0)+1/2 b_{1/2} (\vek',
\vek_0)\}}{(k^2+k^2+\vek \vek'-ME)(k^2-k_0-)}
\frac{d\vek'}{(2\pi)^3}.
\end{array}
\label{9} \ee

Here $a_{3/2}$ is the $Nd$ quartet $(S=3/2)$ scattering amplitude,
while $b_{1/2}$ and $a_{1/2}$ are doublet ($S=\frac12)$ amplitudes
corresponding to the singlet and triplet last $NN$ interaction
respectively. It is seen that the  kernel for  $S=3/2$ is mostly
negative and allows for a faster convergence, in contrast to the
doublet ($S=1/2$) case. Numerical result for quartet scattering
length $a_{3/2}=5.1$fm obtained in \cite{1} is not far from
experimental value \cite{19}, whereas doublet scattering requires
full off-shell calculation \cite{4}.

\section{Hyperspherical Method}

Heretofore the basic dynamics was assumed to be quasi-two-body
(however the  Faddeev technic allows for the full off-shell
description), in  the  sense that typical distance $R$ between an
interacting pair and a third spectator particle is large, $R\gg
r_0$. However this situation is an exclusion, and not the rule,
which can be understood from the representation of the  w.f.
through the 3 body Green's function $(\vexi, \veta$ are Jacobi
coordinates)
\be
\psi (\vexi, \veta)= \psi_0 (\vexi, \veta) + \int G (\vexi-\vexi',
\veta-\veta') V_3 (\vexi', \veta)\psi(\vexi'., \veta') d\vexi'
d\veta'.\label{10}\ee

Here $G(\xi, \veta)=\frac{K_2(\kappa\rho)}{\rho^2}, ~~\kappa=
\sqrt{2m|E|}, ~~ \rho =\sqrt{\vexi^2+\veta^2},$
 and $V_3$ includes all interaction terms. The asymptotics of
 $\psi$ is given by $G$ and is equal to
 \be
 \psi (\vexi, \veta) \sim \frac{1}{\rho^4} ,~~ \rho\to \infty.
 \label{10a}\ee
 Hence the 3-body kynematics tends to concentrate all 3-body w.f.
 inside the interaction region of all 3 particles which generates
 small radius of w.f. even for barely bound 3-body states.  (This
 is also true for $N$ -body systems $N\geq 3$). In this situation
 any pair angular momentum $l_{ij}$ contributes to the total
 energy of the system an amount $\Delta E\sim \frac{l_{ij} (l_{ij}
 +1)}{2m r^2_0}$ which for the 3 nucleon system with $r_0\sim 1$ fm
 and for $l_{ij}=1$ is of the order of $\Delta E\sim 50$ MeV,
 while for the $3q$ system with $m=m_q\sim 0.3$ GeV and $r_0\sim
 0.5$ fm, $\Delta E_q\sim 600 $ MeV.

Therefore it is advantageous to have a wave function with the
minimal number of pair internal angular momenta for the given
total momentum $L$. This basis is provided by hyperspherical
functions ($K$ - harmonics) due to the following properties:

i) the solution of the condition $\hat l_{ij} \Psi =0, i \neq
j=1,... N$ is given by the representation $\Psi_{K=0} = \Psi_0
(\rho)$,
\be
\rho^2= \frac{1}{N} \sum^n_{i<j=1} (\ver_i-\ver_j)^2 \label{11}
\ee where all particles are assumed to have the same mass.

ii) The function $\Psi_K(\ver_i,...\ver_N)=
u_K(\Omega)\chi_K(\rho),$ where $u_K(\Omega)=
\frac{\mathcal{P}_K(\ver_i,...\ver_N)}{\rho^K}$ and $\mathcal{P}_K
$ - harmonic polynomial contains excited angular momenta
$l_1,...l_{N-1}$ the arithmetic sum of which is equal to $K$.

Therefore the basis $\Psi_K$ corresponds to the minimal excitation
of angular momenta and is advantageous for  compact $N$-body
systems. Since as was explained the majority of such systems are
compact, the Hyperspherical Expansion Approach (HEA) \cite{7}
formulated as a system of coupled integral or  differential
equations has proved to be very successful
 both for few-nucleon systems \cite{7},\cite{8}, where short-range
 correlations can be taken into account in the hyperspherical
 correlated basis (last ref. in \cite{18}), and atomic physics
 \cite{9}. It was understood afterwards \cite{10},\cite{11} that
 the HEA works even better for $3q$ systems, since interaction
 there contains no repulsive core and confinement excludes
 two-body channels.

 Therefore already the lowest approximation with $K=0$ yields the
 1\% accuracy for the baryon energy \cite{10,11,12}.

 In this case the baryon state is characterized by the grand
 angular momentum $K$ and radial quantum  number $n=0,1,2$, which
 counts number of zeros  of the w.f. in the $\rho$-space. A
 typical calculation was  done in \cite{12}, and
 the result depends on only two input parameters:  string tension
 $\sigma=0.15$ GeV$^2$ and $ \alpha_s =0.4$, while current masses
 of light quarks have been put to zero. The spin-averaged masses
 $\frac12(M_n+M_\Delta)$ have been computed to eliminate effect of
 hyperfine splitting.\\

To illustrate the simplicity of the method, let us quote the  the
equation for the for the  dominant hyperspherical harmonics
$\psi_K(\rho) =\frac{\chi_K(\rho)}{\sqrt{\rho}},$
\be
-\frac{1}{2\mu} \frac{d^2\psi_K}{d\rho^2} + W_{KK} (\rho)
\psi_K(\rho)= E_K\psi_K(\rho)\label{13}\ee where $W_{KK}(\rho)$
 is the sum of kinetic (angular)
  and potential energies,
  and $\mu$ is a constituent quark mass to be found below dynamically.
  The nonrelativistic appearance of this equation contains nevertheless
  the full relativistic dynamics, since $\mu$ is the einbein field needed
  to get rid of square roots in the relativistic quark action.

  The explicit expression for $W_{KK}$ is
\be
W_{KK} (\rho) =  \frac{d}{2\mu\rho^2} + V_{KK} (\rho),~~
d=(K+\frac32) (K+\frac52), \label{14} \ee while
$V_{KK}(\rho)=(u^+_K(\Omega) \hat V u_K(\Omega))$, is the total
potential $\hat V$, including 2-body and 3-body parts, averaged
over hyperspherical harmonics, which is done analytically. E.g.
for the $Y$-type  $3q$ confining potential one has  $V_{KK}(\rho)
=1.58 \sigma \rho$. It is remarkable that  to find the eigenvalues
$E_K$ with the 1\% accuracy one does not need to solve equation
(\ref{13}), but instead is approximating $W_{KK}(\rho)$ near the
minimum point $\rho_0$ by the oscillator well:
\be
W_{KK}(\rho) = W_{KK}(\rho_0) + \frac12 (\rho-\rho_0)^2
W_{KK}^{\prime\prime}(\rho_0),~~
\frac{dW_{KK}}{d\rho}|_{\rho=\rho_0}=0.\label{15}\ee

The resulting eigenvalues are found immediately: \be E_{Kn} \cong
W_{KK}(\rho_0)+\omega (n+\frac12),~~
\omega^2=W_{KK}^{\prime\prime}/\mu .\label{16}\ee The total baryon
mass is calculated as $M_{Kn}(\mu) =\frac32\mu+ E_{Kn} (\mu)$, and
finally $\mu=\mu_0$ is to be found from the stationary point
condition $\frac{\partial M_{Kn}(\mu)}{\partial\mu}|_{\mu=\mu_0}
=0$. This gives the constituent quark mass
$\mu_0=0.957\sqrt{\sigma} =0.37$ GeV and finally the baryon mass
is $M_{Kn}(\mu_0)$.
 The masses
$\frac12(M_n+M_\Delta)$ computed in this way are shown below in
Table 1.

 \begin{center}
{\bf Table 1}\\

Baryon masses (in GeV) averaged over hyperfine spin splitting
for\\ $\sigma =0.15$ GeV$^2$,  $\alpha_s=0.4,~~ m_i=0$.\\

\begin{tabular}{|l|l|l|l|l|}
\hline State& $M_{Kn}+\lan \Delta H\ran_{self}$& $\lan \Delta
H\ran_{coul}$& $M^{tot}_{Kn}$& $M^{tot}(\exp)$\\\hline $K=0,n=0$&
1.36&-0.274& 1.08&1.08\\ $K=0,n=1$& 2.19&-0.274&1.91& ?\\
$K=0,n=2$&2.9&-0.274&2.62&?\\ $K=L=1,n=0$&1.85&-0.217&1.63&1.6\\
$K=2,n=0$&2.23&-0.186&2.04&?\\\hline
\end{tabular}
\end{center}

As it seen from the Table the calculated spin-averaged mass
$\frac12 (M_N+M_\Delta)$ agrees well with the experimental
average, the same is also true for lowest negative parity states
with $K=L=1$, which should be compared with $\frac{1^-}{2},
\frac{3^{-}}{2}$ states of $N$ and $\Delta$ respectively.

We also notice that breathing modes $(n>0)$ have excitation energy
around 0.8 GeV while orbital excitations $K=L=1$ have energy
interval around 0.5 GeV.

One of important advantages of HEA is that in the lowest
approximation there is no need for numerical computations -- as
demonstrated above
 the result for the mass can be obtained
analytically with 1\% accuracy as can be checked by comparison
with exact calculations, see \cite{10}-\cite{12}.

To conclude, the on-shell approach of STME ( and its Faddeev
generalization) and the HEA are two alternatives  which describe
opposite physical situations. Their coexistence has played a very
important stimulating role for the development of the few-body
physics in the last four decades.

The author is greatly indebted to L.N.Bogdanova for  the help and
useful discussions.

The financial support of RFBR grants 00-02017836, 00-15-96736 and
INTAS grants 00-110 and 00-366 is acknowledged.

\end{document}